\documentclass[superscriptaddress,preprintnumbers,amssymb,nofootinbib]{revtex4-2}

\usepackage{tikz-feynman}
\usepackage{subfig}
\usepackage{bm}
\usepackage{esvect}
\usepackage{verbatim}
\usepackage{multirow}
\usepackage[left]{lineno}
\usepackage{blindtext}
\pdfoutput=1
\usepackage[T1]{fontenc}
\usepackage{graphicx}
\usepackage{epsfig}
\usepackage{amsmath}
\usepackage{amsfonts}
\usepackage{amssymb}
\usepackage{braket}
\usepackage{cancel}
\usepackage{pstricks}
\usepackage{color}
\usepackage{feynmf}
\usepackage[normalem]{ulem}
\usepackage{epstopdf}
\usepackage{soul}
\usepackage{cancel}

\definecolor{ao(english)}{rgb}{0.0, 0.5, 0.0}

\usepackage{tikz}
\usepackage{tikz-feynman}
\tikzfeynmanset{compat=1.1.0}

\usepackage{braket}
\usepackage{cancel}
\usepackage{setspace}
\usepackage{pstricks}
\usepackage{xcolor}
\usepackage{float}
\usepackage{mathtools}
\usepackage{multirow}
\usepackage{booktabs}
\usepackage{enumitem}

\def\missET {{\not\!\! E_T}}

\def\gsim{\lower0.5ex\hbox{$\:\buildrel >\over\sim\:$}}
\def\lsim{\lower0.5ex\hbox{$\:\buildrel <\over\sim\:$}}

\def\be{\begin{equation}}
\def\ee{\end{equation}}
\def\bea{\begin{eqnarray}}
\def\eea{\end{eqnarray}}

\def\op{Q}

\begin{document}

\title{Counting b-jets at the FCC-ee as a probe of top-quark flavor physics}

\author{Shaouly Bar-Shalom}
\email{shaouly@physics.technion.ac.il, shaouly@gmail.com}
\affiliation{Physics Department, Technion--Institute of Technology, Haifa 3200003, Israel}
\author{Jose Wudka}
\email{jose.wudka@ucr.edu}
\affiliation{Physics Department, University of California, Riverside, CA 92521, USA}

\date{\today}

\begin{abstract}
We explore the sensitivity of a future FCC-ee to multi-TeV new physics (NP) that can generate vector, scalar and tensor 
flavor-changing (FC) $eetu_k$ contact interactions ($u_k=u,c$ for $k=1,2$), probed via single top-quark production, $e^+ e^- \to t \bar{u}_k, t \bar{u}_k \gamma$ (+ c.c.) and leading (following the top-quark decay) to two- and four-jet signals, $e^+ e^- \to 2j + X, 4j + X$, where $X$ denotes any non-jet final-state particles.
Our analysis exploits an approximately conserved Standard Model (SM) quantum number introduced in \cite{our_bP_paper} and termed "$b$-Parity" ($b_P$), which is applicable to scattering processes  of the type $e^+ e^- \to n \cdot j_b + m \cdot j_\ell + X$, where $n$ and $m$ are the number of produced $b$-jets and 
light-jets ($u,d,c,s,$ and/or gluons) and $b_P=(-1)^n$. 
The FC $eetu_k$ four-fermion interactions can generate distinct $b_P$-odd ($b_P = -1$) signals in these multi-jet events, for which the only significant SM background stems from $b$-jet misidentification. 
We demonstrate that the FCC-ee, operating at a $\sqrt{s} = 240$ GeV (the $ZH$ run phase) with high luminosity and excellent flavor-tagging performance, is an ideal platform to search for these $b_P$-odd signatures.
%
Indeed, by simply counting the number of final-state $b$-jets, the FCC-ee can can probe NP scales of $\Lambda \sim 10$ TeV for the new heavy states that generate the vector, scalar and tensor $eetu$ and/or $eetc$ interactions.
This reach, remarkably about 40 times the assumed FCC-ee center-of-mass energy, improves upon current bounds on some of these four-fermion operators by an order of magnitude and it critically relies on high-purity b-jet tagging. 
\end{abstract}

\maketitle
\flushbottom

\newpage 

\section{Introduction \label{sec:intro}}

The Standard Model (SM) of particle physics is understood to be a low-energy effective field theory (EFT) rather than a full description of the microscopic nature, despite its remarkable successes over the past decades. Therefore, the search for a more fundamental framework is a primary focus of the ongoing experiments at the LHC and is also the central driver for next-generation particle colliders.
One promising candidate for a future collider, which will be of our focus in this work, is the Future Circular $e^+e^-$ Collider (FCC-ee), which is planned to run at center-of-mass (CM) energies starting from  the $Z$-pole and up to top-quark pair production threshold and delivering very high luminosities in each phase \cite{Blondel:2021ema,FCCex_report_2025,FCC:2025uan,FCC:2025jtd}. Consequently, the FCC-ee offers a high-precision platform for indirect searches of heavy new physics (NP) beyond the SM (BSM), which provides a unique testing ground for rare phenomena like flavor-changing (FC) processes that may originate at the multi-TeV scale.

In this paper, we explore the potential of the FCC-ee to probe TeV-scale FC top-quark interactions, the detection of which would provide unambiguous evidence for BSM physics.
Indeed, due to the significantly larger top-quark mass compared to all other quarks, 
FC top-quark interactions best manifests the SM flavor problem and makes it the most sensitive to several types of NP; in particular, to new flavor and CP-violating physics~\cite{ourreview,0409342}. 
We recall that tree-level Flavor-Changing Neutral Currents (FCNC) are absent in the SM and are GIM suppressed at the loop level; this is the case for
example in $t \to u$ and $t \to c$ transitions in top decays~\cite{FCtopdecay1,FCtopdecay2,FCtopdecay3,FCtopdecay4,FCtopdecay5,FCtopdecay6,FCtopdecay7,0102037,FCtopdecay8,t_to_cdecay_soni,t_to_c_Hou} and/or in top-quark production processes~\cite{FCtopprod1,FCtopprod2,FCtopprod3,FCtopprod4,FCtopprod5,eetc_soni1,eetc_soni2,eetc_Hou1}. 
Thus, the feeblest signal of FCNC effects in the top-quark sector, either direct or indirect, may be an indicator of new BSM physics. This fact has led to a considerable theoretical effort as well as experimental activity in understanding and searching for top FCNC within model independent approaches, as well as within specific popular BSM models.

In this work we focus on FC  single top-quark production processes at the FCC-ee, running at $E_{CM}=240$ GeV (the $ZH$ phase) and an expected total integrated luminosity (over 3 years) of about 10 ab$^{-1}$ \cite{FCCex_report_2025,FCC:2025uan,FCC:2025jtd} (the initial planned phases of the FCC-ee, operating at CM energies around the $Z$-pole and the $WW$ threshold, lack the necessary energy to study single top-quark production). Indeed, single top-quark production at the LHC and at future colliders provides a promising window into the top-quark sector for BSM physics searches \cite{our_tc_paper,Gudron1,Gudron2,2012.05735,dEnterria:2016fpc,Degrande:2018fog,Maltoni-global,Hartland:2019bjb,Durieux:2019rbz,singletop1,tZgamma1,SMEFTtop1,tZ3,SMEFTtop2,SMEFTtop3,SMEFTtop4,Zhang:2016omx,tH1,tH2,tH3,tH4,tH5,tH6,tH7,tH8,tZ1,tZ2,tZ4,tZ5,tZ6,tZ7,tZ8-trilepton,tZ9-trilepton,tgamma1,tZEFT-decay1,tZprime,1906.04573,1906.04884,our_LFU_paper,Afik:2021jjh,Jose_thj,our_ttll_paper,2302.01143,ourPRL,Jueid:2024nbv,our_EIC_paper,Corcella:2025idg,2604.13562}, since it can, 
in some cases, partly bypass the $1/\Lambda$ suppression expected from the heavy NP scale ($\Lambda$). For recent studies on flavor physics at the FCC-ee 
see e.g., \cite{Corcella:2025idg,2207.04844,2306.17520,Greljo1,2503.17019,2510.23488}.
 
We use the so-called Standard Model Effective Field Theory (SMEFT) to describe the underlying FC interactions in the top-quark sector; the SMEFT framework provides a systematic setup for parameterizing BSM physics in terms of higher-dimensional operators built from SM fields and obeying its local symmetries, thus enabling a model-independent study of physics at  scales well beyond the CM energies of present and future colliders.
Flavor physics is an integral and important component of the SMEFT framework, which, as will be shown in this work, offers very sensitive probes of possible new heavy  states, even when their mass lays at the multi TeV-scale, i.e., far above the electroweak and the FCC-ee energy scale.
Global and comprehensive EFT studies of various types of higher dimensional operators involving the top-quark field(s) can be found in~\cite{1008.3562,Maltoni-global,Hartland:2019bjb,Durieux:2019rbz,SMEFTtop1,tZ3,SMEFTtop2,SMEFTtop3,Gudron2,SMEFTtop4,Ellis:2020unq,Ethier:2021bye,2503.11518}, where studies of the effects of (2-quarks)(2-leptons) 4-Fermi operators (which are of interest in this study) have been conducted in~\cite{Gudron1,Gudron2,bsll,bbll-our,our_LFU_paper,Afik:2021jjh,Tonero:2020zcy,Maltoni-global,topdecay1,1008.3562,topdecay3,topdecay2,Gottardo:2019lmv,our_ttll_paper,our_EIC_paper}; note that the FC $\ell \ell tu, \ell \ell tc$ class of operators are poorly bounded as will be further discussed below.

In what follows, we will adopt the approach originally introduced by the authors more than two decades ago in \cite{our_bP_paper}; to test FC BSM effects in 3rd generation quark interactions by simply counting the number of $b$-jets in the final state (FS) of multi-jet production channels. As outlined in \cite{our_bP_paper}, this approach is best suited for lepton colliders (such as the FCC-ee) but, in some cases, it may be extended to hadron colliders and/or electron-proton colliders if the $b$-quark content in the proton can be ignored, e.g., in  single and di-jet production at the future Electron Ion Collider (EIC), as we have recently suggested in \cite{our_EIC_paper}.
We will then show that, by simply counting the number of $b$-quark jets in the FS, the FCC-ee will be sensitive to new heavy states which generate 3rd generation FC interactions at scales beyond  $\Lambda \sim 10$ TeV. 

\section{$b$-Parity in multi-jet production at the FCC-ee  \label{sec:bp}}

The physics behind the method we use below is based on the observation that in the limit where the off-diagonal 3rd generation CKM entries vanish, i.e., $V_{3j},V_{j3}=0$ ($j\neq 3$), the SM acquires an additional global $U(1)_b$ symmetry ("bottomness"), where $b$ and $\bar b$ have opposite charges, and which holds to any order in perturbation theory. Since $V_{j3},V_{3j}$ ($j \neq 3$), though not zero, are small (recall that $ V_{31},V_{13} \sim {\cal O}(\lambda^3)$ and $ V_{32},V_{23} \sim {\cal O}(\lambda^2)$, where $\lambda \sim 0.22$ is the Wolfenstein parameter \cite{Wolfenstein:1983yz}), the quantum number associated with this $U(1)_b$ is approximately conserved (in particular, this implies that the top quark decays almost exclusively into $ bW$). Then, given a reaction of the type
\begin{eqnarray}
    n_i \cdot b + X \to n_f \cdot b + Y ~, \label{bp1}
\end{eqnarray}
where 
\begin{enumerate}[label=\roman*.]
\item $X,Y$ denote sets of particles not containing  $b$-quark/jets,
\item $n_i,n_f$ are the net numbers of $b$-quark/jets in the initial and final states, respectively, 
\item there are no top quarks in the initial state, and those in the final state have decayed via $t \to bW$,
\end{enumerate}
then to an excellent approximation we will have  $(-1)^{n_i} = (-1)^{n_f}$, so that we can define an approximately conserved flavor number for such collider scattering processes, $b_P \equiv (-1)^{n_f-n_i}$, which was named $b$-Parity in \cite{our_bP_paper}.
In particular, since the initial state in $e^+e^-$ colliders contains no $b$-quarks ($n_i=0$ in Eq.~\ref{bp1}), an odd number of $b$-quark jets {\it will not} be generated in the SM in scattering processes in such colliders, as was pointed out in \cite{our_bP_paper}.

We will thus apply the concept of $b$-Parity for searches of new BSM  FC physics in the top-quark sector in multi-jet production at the FCC-ee:
\begin{eqnarray}
e^+  e^- \to n \cdot j_b + m \cdot j_c + \ell \cdot j_\ell + X \label{nbx}~,
\end{eqnarray}
where $n$ denotes the number of $b$ and/or $\bar b$ jets\footnote{The method used here does not require differentiating between $b$ and $\bar b$-quark jets.} in the FS, $m$ is the number of c-quark jets, $\ell$ the number of light-quark and/or gluon jets and $X$ stands for leptons, photons and/or missing energy. For the reaction in Eq.~\ref{nbx}, the $b$-Parity number becomes: 
\begin{eqnarray}
b_P=(-1)^n ~, \label{bParity}
\end{eqnarray}
so that the measured quantum number reduces to the net number of detected $b$-quark jets in the final state, in which case it is convenient to use the derived quantity $ b_P $.

As mentioned earlier, we will focus below on FC single top-quark production 
in an FCC-ee running phase with a CM energy of $E_{CM} =240$ GeV. 
The leading SM processes that violate $b_P$ in these multi-jet production processes at the FCC-ee necessarily involve the charge-current $u \to b$ and/or $c \to b$ interactions, which are suppressed by the corresponding small off-diagonal CKM element factors, $|V_{ub}|^2$ and $|V_{cb}|^2$, so that the SM is $b_P$-even at the FCC-ee to very good accuracy, as will be further discussed in section \ref{sec:FCC-ee240}.
Thus, to experimentally detect $b_P$-odd ($b_P=-1$)
NP signals in the processes of the type shown in Eq.~\ref{nbx}, one should simply measure/count the number of events with an odd number of $b$-jets in the FS; this 
presents a jet-reconstruction challenge and
will require jet flavor-tagging performance at the FCC-ee to be sufficiently high; 
in particular, a satisfactory $b$-jet tagging efficiency and high purity of the $b$-jet samples. 
Indeed, considerable attention has already been devoted to jet flavor-tagging at the future FCC-ee collider \cite{Bedeschi:2022rnj,Gautam:2022szi,Blekman:2024wyf,Higgs-physics-jet-tagging,Beck:2025xnu} and its importance for Higgs and top-quark physics as well as for precision and NP studies at the FCC-ee has been recently highlighted \cite{Azzi:2021gwg,2306.17520,Greljo1,Higgs-physics-jet-tagging}.

The important tagger parameters for our study are the $b$-jet tagging efficiency
$\epsilon_b$ (true positive) and the purity factors $t_c$ and $t_j$, which are the (false positive) probabilities for mistagging a $c$-quark jet and light-quark and/or gluon jet, respectively, for a $b$-jet (we refer to the $u,d$ and $s$ quarks as light-quarks).\footnote{We have separated the false positive purity factors of $c$-quark jets and the other light-quarks and gluon jets, since in general one expects $t_c > t_j$ \cite{Bedeschi:2022rnj,Blekman:2024wyf}. We have also unified the false positive gluon and light quarks jets factors, which could in principle be different \cite{Bedeschi:2022rnj,Blekman:2024wyf}, since the SM background with gluons in the FS (in the $e^+ e^- \to 4j, 4j \gamma$ channels, see next section) is sub-leading compared to the four-quark FS for these processes.} In particular, $1-\epsilon_b$ corresponds to the (false negative) probability of misidentifying a $b$-jet as a light or $c$-quark jet and $1-t_c$, $1-t_j$ correspond to the (true negative) probabilities of identifying a non $b$-quark jet as a light-quark or $c$-quark jet. In general, tagger algorithms are designed to minimize $t_c$ and $t_j$ thereby improving purity (for a fixed b-jet tagging efficiency $\epsilon_b$); typically the higher the $b$-tagging efficiency is the lower the purity. Accordingly, following the analysis in \cite{Bedeschi:2022rnj,Gautam:2022szi,Blekman:2024wyf,Higgs-physics-jet-tagging,Beck:2025xnu}, we will consider below three benchmark $b$-jet tagger scenarios, "loose", "medium" and "tight"; in particular, assigning a lower purity (i.e., higher values of $t_c,t_j$) for higher values of $b$-tagging efficiency ($\epsilon_b$):
%
\begin{equation*}
   \begin{array}{|c||c|c|c|}
    \hline
    \text{tagger} & ~~\epsilon_b~~ & ~~t_c~~ & t_j \cr
    \hline
    {\bf loose} & 0.9 & 0.01 & 0.001 \cr
    \hline
    {\bf medium} & 0.8 & 0.005 & 0.001 \cr
    \hline
    {\bf tight} & 0.7 & 0.001 & 0.0001 \cr
    \hline
    \end{array}   
\end{equation*}
and in some instances we will fix $\epsilon_b,t_c$ to their values below and very the light-quark purity factor in the range $0.0001 < t_j < 0.01$.

Using the tagger efficiency parameters defined above, we can write the probability (or cross-section) for detecting precisely $k$ $b$-jets in the reaction of Eq.\ref{nbx} is given by
\begin{eqnarray}
\bar\sigma_{kj_b} = \sum_{u,v,w} P_u^n P_v^m P_w^\ell
\left[ \epsilon_b^u (1-\epsilon_b)^{n-u} \right] 
\left[ t_c^v (1-t_c)^{m-v} \right]
\left[ t_j^w (1-t_j)^{\ell-w} \right] \sigma_{ n m \ell} ~ \delta_{u+v+w,k} 
\label{sigmabar} ~,
\end{eqnarray}
where $P^i_j = i!/j!/(i-j)!$ and 
\begin{eqnarray}
    \sigma_{ n m \ell} = \sigma (e^+ e^- \to n \cdot j_b + m \cdot j_c + \ell \cdot j_\ell + X) ~. \label{nml}
\end{eqnarray}

The main obstacle in the search for $b_P$-odd NP signals is, therefore, the reducible SM background due to jet misidentification, which results from having a non-optimal $b$-tagging efficiency ($\epsilon_b < 1 $), and/or having non-zero false positive probabilities ($t_c$, $t_j$) of mistagging a non $b$-jet for a $b$-jet (i.e., non-perfect purity of the $b$-jet sample). This type of background would of course disappear as $\epsilon_b \to 1$ and $t_c, t_j \to 0$, but even the seemingly high purity of e.g., $t_j \sim 0.001$ and/or $t_c \sim 0.01$ and a relatively high $b$-tagging efficiency of $\epsilon_b \sim 0.8$, can produce a significant number of  "fake" (misidentified) $b_P$-odd events in the detector, as will be shown below. 

\section{EFT description of top-quark flavor changing physcis at the FCC-ee  \label{sec:3rdflavor}}

In the SMEFT framework the underlying NP effects are parameterized by higher dimensional, gauge-invariant effective operators, $\op_i^{(n)}$ (where $n$ denotes the mass dimension), where the effective operators are constructed using the SM fields and their coefficients are suppressed by inverse powers of the NP scale $\Lambda$ \cite{EFT1,EFT2,EFT3,EFT4,EFT5}:
\begin{eqnarray}
{\cal L} = {\cal L}_{SM} + \sum_{n=5}^\infty
\frac{1}{\Lambda^{n-4}} \sum_i \alpha_i \op_i^{(n)} \label{eq:EFT1}~,
\end{eqnarray}
assuming that the heavy NP is decoupling and weakly-coupled. One can further divide the higher-dimension effective operators into those that can be potentially generated at tree-level (PTG) and those that are necessarily loop generated (LG) in the underlying heavy theory \cite{jose_PTG_LG}. The dominating NP effects are then expected to be generated by the contributing PTG operators with the lowest dimension (smallest $n$). The (Wilson) coefficients $\alpha_i$ depend on the details and dynamics of the underlying NP and, therefore,  they parameterize all possible weakly-interacting and decoupling types of underlying interactions of the heavy degrees of freedom. In particular, it is expected that  $\alpha_i =O(1)$ for the PTG operators (for favorable types of NP), and  $\alpha_i \sim 1/(4\pi)^2$ for all LG operators; the effects of LG operators are thus a-priori suppressed by a loop factor and, therefore, their effects at lower energies (i.e., $E < \Lambda$) are expected to be subleading. 

For the calculations below we will generally consider the effects of a single operator, for which we will use a more practical definition of the "effective" NP scale, which is the variable that enters the calculation: 
\begin{eqnarray}
    \Lambda_{\tt eff} \equiv \frac{\Lambda}{\sqrt{\alpha}} ~; \label{lam_eff}
\end{eqnarray}
%
note that $\Lambda_{\tt eff} \sim \Lambda$ for natural PTG couplings. 

%

\begin{figure}[htb]
  \centering
\begin{tikzpicture}
  \begin{feynman}
  
    \vertex (eplus)  at (-2,  1) {$e^{+}$};
    \vertex (eminus) at (-2, -1) {$e^{-}$};
    \vertex (v)      at ( 0,  0);
    \vertex (top)    at ( 2,  1) {$t$};
    \vertex (jet)    at ( 2, -1) {$j$};

    \diagram*{
      (eminus) -- [fermion] (v),        
      (v) -- [anti fermion] (eplus),    
      (v) -- [fermion] (top),           
      (v) -- [fermion] (jet),           
    };

    \filldraw[black] (v) circle (3pt);


\vertex (eplus2)  at (3,  1.5) {$e^{+}$};
\vertex (eminus2) at (3, -1.5) {$e^{-}$};

\vertex (v2a) at (5,  0.8);
\vertex (v2b) at (5, -0.8);

\vertex (top2) at (7,  1.5) {$t$};
\vertex (jet2) at (7.8, -1.5) {$V=\gamma,\,Z,\,W$};
\vertex (V2)   at (7, 0) {$j$};

\diagram*{
  (eminus2) -- [fermion] (v2b) -- [fermion] (v2a) -- [anti fermion] (eplus2) ,

  (v2a) -- [fermion] (top2),
  (v2b) -- [photon] (jet2),

  (v2a) -- [fermion] (V2),
};

\filldraw[black] (v2a) circle (3pt);

\end{feynman}
\end{tikzpicture}
\caption{Representative Feynman diagrams for 
$e^+ e^- \to t j$ and $e^+ e^- \to t j + V$, where $V=\gamma,Z$ or $W$.}
  \label{fig:1}
\end{figure}
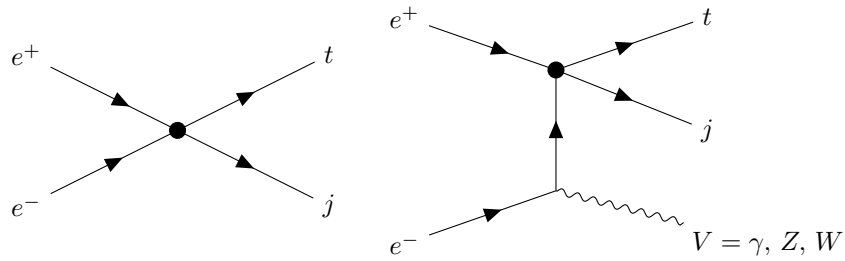

We will examine below new flavor structures in top-quark physics that can be tested in  
single-top production at the FCC-ee via (see Fig.~\ref{fig:1}):
\begin{eqnarray}
    && e^+ e^- \to t j + \text{\tt c.c.} \label{Eq:tj} ~, \\
    && e^+ e^- \to t j V +  \text{\tt c.c.}  ~ (V=\gamma,Z,W)  \label{Eq:tjV} ~,
\end{eqnarray}
where \text{c.c.} denotes the charge conjugate FS and, for the NP we consider below, $j$ stands specifically for a light-quark jet originating from a $u$ or $c$ quark. As detailed in Table \ref{tab:3} below, these single top production processes lead  to various types of $2,4$ and 6 jets signals. 
\begin{table*}[htb]
\caption{
The potential multi-jets signals corresponding to the single top production processes $e^+ e^- \to tj,~ tj V$ (+ c.c.) following the top-quark decays, where $V=\gamma,Z,W$. The single top signals result from 
the dim.6 FC 4-fermi effective operators of Table \ref{tab:2}. Also, $\ell$ is a charged lepton ($\ell=e^\pm,\mu^\pm$ or $\tau^\pm$) and $\missET$ is missing transverse energy.
\label{tab:3}}
\renewcommand{\arraystretch}{1.5}
\begin{tabular}[t]{c|c|c}
  & $t \to bW(\to jj)$ & $t \to bW(\to \ell \nu)$ \\
\hline
\hline 
$e^+ e^- \to t j$ & $4j$ ($1b$) & $2j+\ell+\missET$ ($1b$) \\
$e^+ e^- \to t j \gamma$ & $4j + \gamma$ ($1b$) & $2j+\gamma+ \ell+\missET$ ($1b$)  \\
$e^+ e^- \to t j Z( \to q \bar q,~ q \neq b)$ & $6j$ ($1b$) & $4j+\ell+\missET$ ($1b$) \\
$e^+ e^- \to t j Z( \to b \bar b)$ & $6j$ ($3b$) & $4j+\ell+\missET$ ($3b$)  \\
$e^+ e^- \to t j Z( \to \ell^+ \ell^-)$ & $4j + 2\ell$ ($1b$) & $2j+3\ell+\missET$ ($1b$) \\
$e^+ e^- \to t j Z( \to \nu \bar\nu)$ & $4j + \missET$ ($1b$) & $2j+\ell+\missET$ ($1b$) \\
$e^+ e^- \to t j  W(\to q \bar q^\prime, q,q^\prime \neq b)$ & $6j$ ($1b$) & $4j+\ell+\missET$ ($1b$) \\
$e^+ e^- \to t j W(\to \ell^- \bar\nu)$ & $4j+\ell+\missET$ ($1b$) &  $2j+2\ell+\missET$ ($1b$) \\
\end{tabular}
\end{table*}

The PTG operators of dimension six (dim.6) that can generate these signals are listed in Tables \ref{tab:1} and \ref{tab:2}. 
Note in particular,
\begin{itemize}
\item The operators $\op_{\ell e dq}$, $\op_{\ell e qu}^{(1)}$ and $\op_{\ell e qu}^{(3)}$ are non-hermitian and no symmetry applies to their indices. Thus, for example, their $(113k)$ entry is different from the $(11k3)$ one ($k \neq 3$).\footnote{The numbers denote the generation of the fermions as they appear in the operator label.} We note that, for the charged current  reactions $e e \to t j W$ (see Fig.~\ref{fig:1}) only the $(11k3)$ entries of these operators contribute.
\item The operators $\op_{\ell q}^{(1)}, \op_{\ell q}^{(3)},  \op_{\ell u},\op_{\ell d},\op_{qe},\op_{\ell u},\op_{\ell d}$ are all hermitian and are composed of two hermitian currents, so that their $(113k)$ and $(11k3)$ entries are the same, e.g., $\op_{\ell q}^{(1)}(113k) = \op_{\ell q}^{(1)}(11k3)$ etc.
\item The operators $\op_{\ell q}^{(1)}(113k)$, $\op_{\ell q}^{(3)}(113k)$, $\op_{qe}(113k)$ and $\op_{\ell e q d}(11k3)$ 
also generate, due to SU(2) invariance, the FC $eebd$ ($k=1$) and $eebs$ ($k=2$) contact terms, which can also mediate the $b_P$-odd single $b$-jet signals $e^+ e^- \to bj,bjV$ ($V=\gamma,Z$). These are included in our calculations, but their contribution to the single $b$-jet NP signals of interest are much smaller. Note that $\op_{\ell e q d}(113k)$ which also generates the FC $eebd$ ($k=1$) and $eebs$ ($k=2$) sclar interactions, does not contribute to the single top production process that we consider and is, therefore, not included in our calculations.  
\item The operators $\op_{\ell q}^{(3)}(113k)$ and $\op_{\ell e qu}^{(1)}(113k)$ and $\op_{\ell e qu}^{(3)}(113k)$ also generate, due to SU(2) invariance, the FC $e \nu bu$ ($k=1$) and $e \nu bc$ ($k=2$) charged current contact terms, which can mediate the $b_P$-odd single $b$-jet signal $e^+ e^- \to bjW$. The effect of this channel on our $4j +X $ and $2j +X$ signals is much smaller and can be neglected, though it is included in our calculation. 
\end{itemize}

\begin{table*}[htb]
\caption{\label{tab:1}
The dim.6 PTG 4-fermion operators in the SMEFT, which potentially involve FC top quark interactions with leptons.
The operators are ordered according to their type: vector, scalar and tensor and their helicity structure. The subscripts $p,r,s,t$ are flavor indices.}
\begin{center}
\small
\begin{minipage}[t]{4.45cm}
\renewcommand{\arraystretch}{1.5}
\begin{tabular}[t]{c|c}
\multicolumn{2}{c}{vector: $(\bar LL)(\bar LL)$} \\
\hline
${\cal O}_{lq}^{(1)}(prst)$                & $(\bar l_p \gamma_\mu l_r)(\bar q_s \gamma^\mu q_t)$ \\
${\cal O}_{lq}^{(3)}(prst)$                & $(\bar l_p \gamma_\mu \tau^I l_r)(\bar q_s \gamma^\mu \tau^I q_t)$
\end{tabular}
\end{minipage}
\begin{minipage}[t]{4.45cm}
\renewcommand{\arraystretch}{1.5}
\begin{tabular}[t]{c|c}
\multicolumn{2}{c}{vector: $(\bar RR)(\bar RR)$} \\
\hline
${\cal O}_{eu}(prst)$                      & $(\bar e_p \gamma_\mu e_r)(\bar u_s \gamma^\mu u_t)$ \\
%
\end{tabular}
\end{minipage}
\begin{minipage}[t]{4.45cm}
\renewcommand{\arraystretch}{1.5}
\begin{tabular}[t]{c|c}
\multicolumn{2}{c}{vector: $(\bar LL)(\bar RR)$} \\
\hline
${\cal O}_{lu}(prst)$               & $(\bar l_p \gamma_\mu l_r)(\bar u_s \gamma^\mu u_t)$ \\
%
${\cal O}_{qe}(prst)$               & $(\bar e_p \gamma^\mu e_r)(\bar q_s \gamma_\mu q_t)$ \\
\end{tabular}
\end{minipage}

\vspace{0.25cm}

\begin{minipage}[t]{5.45cm}
\renewcommand{\arraystretch}{1.5}
\begin{tabular}[t]{c|c}
\multicolumn{2}{c}{scalar: $(\bar LR)(\bar L R) ~\&~ (\bar LR)(\bar R L) +\hbox{h.c.}$} \\
\hline
${\cal O}_{lequ}^{(1)}(prst)$ & $(\bar l_p^j e_r) \epsilon_{jk} (\bar q_s^k u_t)$ \\
${\cal O}_{ledq}(prst)$ & $(\bar l_p e_r) (\bar d_s q_t)$
\end{tabular}
\end{minipage}
\begin{minipage}[t]{5.45cm}
\renewcommand{\arraystretch}{1.5}
\begin{tabular}[t]{c|c}
\multicolumn{2}{c}{tensor: $(\bar LR)(\bar L R) +\hbox{h.c.}$} \\
\hline
${\cal O}_{lequ}^{(3)}(prst)$ & $(\bar l_p^j \sigma_{\mu\nu} e_r) \epsilon_{jk} (\bar q_s^k \sigma^{\mu\nu} u_t)$ \\
\end{tabular}
\end{minipage}
\end{center}
\end{table*}


\begin{table*}[htb]
\caption{
The dim.6 effective operators that are PTG in the underlying heavy theory and that can mediate the single-top production signals $e^+ e^- \to t j, tj V$ ($V=\gamma, Z,W$  and $j$ is a light-quark jet), through the  FC $eetu_k$ and $e \nu t d_k$ 4-fermion interactions ($k \neq 3$), as indicated. 
\label{tab:2}}
\renewcommand{\arraystretch}{1.5}
\begin{tabular}[t]{c|c|c}
 & $e^+ e^- \to t j, tj\gamma, tjZ$ &  $e^+ e^- \to t j W$  \\
   & ($e et u_k$) &  ($e \nu t d_k$)  \\
\hline
$\op_{\ell q}^{(1)}(113k)$ & $\checkmark$  & $\times$ \\
$\op_{\ell q}^{(3)}(113k)$ & $\checkmark$ & $\checkmark$ \\
$\op_{eu}(113k)$ & $\checkmark$ &  $\times$ \\
$\op_{ \ell u}(113k)$ & $\checkmark$ & $\times$ \\
$\op_{q e}(113k)$ & $\checkmark$ & $\times$ \\
$\op_{\ell e  dq }(11k3)$& $\times$   &  $\checkmark$ \\
$\op_{\ell e q u}^{(1)}(113k)$ &  $\checkmark$  &  $\times$ \\
$\op_{\ell e q u}^{(1)}(11k3)$ &  $\checkmark$  &  $\checkmark$  \\
$\op_{\ell e q u}^{(3)}(113k)$ &  $\checkmark$  &  $\times$ \\
$\op_{\ell e q u}^{(3)}(11k3)$ &  $\checkmark$  &  $\checkmark$  \\

\end{tabular}
\end{table*}

Let us briefly describe the current bounds on the scale of the operators listed in Table \ref{tab:2}, which are relevant for the single top production signals of interest here. These were studied and constrained directly from single top-quark production $e^+ e^- \to t \bar u, t \bar c$ (+ c.c.) at LEP2 \cite{our_tc_paper,L3,DELPHI}, where a lower limit of $\Lambda_{\tt eff} \gsim 0.5 - 1.5$ TeV was found, depending on the underlying NP mechanism; $\Lambda_{\tt eff} \gsim 0.5$ TeV for the scalar $eetu/eetc$ operators and $\Lambda_{\tt eff} \gsim 1.5$ TeV for a tensor-like $eetu/eetc$ 4-Fermi vertex.  
A slight improvement can be obtained by combining these LEP2 bounds with (the rather weak) bounds derived from the rare top decay to a pair of charged leptons and a jet $t \to j \ell^+ \ell^-$  \cite{Maltoni-global,topdecay1,1008.3562,topdecay3,topdecay2}. Furthermore, a recent study \cite{Afik:2021jjh} of the effects of the tensor(vector) $eetu$ 4-Fermi operators in single top + dilepton events at the LHC, i.e., $pp \to t \ell^+ \ell^-$, have found that a 95\% CL bounds of $\Lambda_{\tt eff} \gsim 5(3)$ TeV is expected with $\sim {\cal O}(100)$ fb$^{-1}$ of data if a dedicated analyses will be performed for this purpose. A higher sensitivity to these tensor and vector $ee tu$ and $ee tc$ contact interactions is expected at the HL-LHC, reaching e.g., $\Lambda_{\tt eff} \gsim 7 $ TeV for the $\ell \ell tu$ tensor term via $pp \to t \ell \ell$ ($\ell=e,\mu$) \cite{Afik:2021jjh} and $\Lambda_{\tt eff} \gsim 5 $ TeV for the $\ell \ell tc$ tensor operator in the $t \bar t$ production channel, $pp \to t \bar t$, followed by the top decays $t \to c \ell \ell$ \cite{2502.18667}.  

Tighter bounds can be obtained on the 4-fermion operators in Table \ref{tab:2} that can also generate the FC $eebd,eebs,e \nu bu$ and $e \nu bc$ contact interactions, $\op_{\ell q}^{(1)}(113k)$, $\op_{\ell q}^{(3)}(113k)$, $\op_{qe}(113k)$, $\op_{\ell e q d}(11k3)$, $\op_{\ell e qu}^{(1)}(113k)$ and $\op_{\ell e qu}^{(3)}(113k)$ (see discussion above). In particular, these operators effect $B$-decays and also Drell-Yan (DY) processes at the LHC, which are initiated by the $b$-quark content in the protons, i.e., from $\bar b u \to e^+ \nu_e$ and  $\bar b  d \to e^+  e^-$; bounds derived from the DY interactions at the LHC yield $\Lambda_{\tt eff} \gsim 3-7$ TeV, depending on the type of operator and process \cite{Greljo:2022jac,Allwicher:2022gkm,Greljo:2023bab,Grunwald:2023nli,Hiller:2025hpf}, while bounds obtained from the $b$-decays $b \to u e \nu_e$ and $b \to d ee$ are considerably tighter, reaching in some cases 
$\Lambda_{\tt eff} > {\cal O}(10-20)$ TeV when a single NP operator is turned on \cite{Greljo:2022jac,Greljo:2023bab}.

Finally, interesting indirect bounds on the $eetu$ and $eetc$ 4-Fermi operators can be applied under certain conditions, by relating existing constraints from flavor-conserving processes to NP contributions to the flavor-violating top decay $t \to j \ell^+ \ell^-$ \cite{2303.00781}. 

We will analyze below the effects of the scalar, vector and tensor operators in Tables \ref{tab:1} and \ref{tab:2}, "turning on" a single operator at a time for each category. 
Specifically, we find that the sensitivity depends on the operator class: scalar, vector, or tensor, but remains the same for operators within the same class, since the single top cross-sections in Eqs.~\ref{Eq:tj} and \ref{Eq:tjV} are the same within a given class of operators. For example, 
the cross-section for the single top-quark production process $e^+ e^- \to t \bar u_k + \bar t u_k$ in Eqs.~\ref{Eq:tj} ($u_k=u,c$ for $k=1,2$, respectively), generated by of the scalar, vector or tensor $eetu_k$ 4-Fermi interactions is \cite{our_tc_paper}:
\begin{eqnarray}
    \sigma(e^+ e^- \to t \bar u_k + \bar t u_k) = \frac{s}{\Lambda_{\tt eff}^4} \frac{\beta_t^2}{4 \pi (1+\beta_t)^3} {\cal F}_{S,V,T} ~, \label{CSXtu}
\end{eqnarray}
where $\beta_t=(s-m_t^2)/(s+m_t^2)$ and ${\cal F}_{S,V,T}$ correspond to the contributions from the scalar, vector and tensor operators, respectively, given by 
\begin{eqnarray}
    {\cal F}_{S} = \frac{3}{2} (1+\beta_t)~,~ {\cal F}_{V} = (3+\beta_t) ~,~{\cal F}_{T} = 8(3-\beta_t) ~,  
\end{eqnarray}
so that ${\cal F}_{S}$ applies to $\op_{\ell e q u}^{(1)}(113k)$ or $\op_{\ell e q u}^{(1)}(11k3)$, ${\cal F}_{V}$ applies to 
$\op_{\ell q}^{(1)}(113k)$, $\op_{\ell q}^{(3)}(113k)$, $\op_{eu}(113k)$, $\op_{\ell u}(113k)$ or $\op_{qe}(113k)$ and ${\cal F}_{T}$
applies to $\op_{\ell e q u}^{(3)}(113k)$ or $\op_{\ell e q u}^{(3)}(11k3)$.

Consequently, the analyses presented below for a scalar, vector, and tensor operators holds for any of the operators in Table \ref{tab:2} in their respective categories. 

\begin{table*}[htb]
\caption{
Quantum numbers of the NP bosons $X$ that can generate the operators in Table \ref{tab:1} at tree level. $Y$ denotes the hypercharge; the numbers under $SU(2)$ give the dimension of the corresponding irreducible representation; and spin $1$ and $1'$ correspond, respectively, to the representations $(1/2,\,1/2) $ and $ (1,0) \oplus (0,1)$ of the Lorentz group.
\label{tab:NP}}
\renewcommand{\arraystretch}{1.5}
\begin{tabular}[t]{c|c|c|c||c|c|c|c|c|c|c|c|}
spin & $Y$ & $SU(2)$ & $SU(3)$ & $\op_{lq}^{(1)}$ & $\op_{lq}^{(3)}$ & $\op_{eu}$ & $\op_{lu}$ & $\op_{qe}$ & $\op_{lequ}^{(1)}$ & $\op_{leqd}$ & $\op_{lequ}^{(3)} $ \cr \hline
\multirow{5}{1em}{$0$}  & $1/3$ &  $1$ & $\bar3$ & $\checkmark$ & $\checkmark$ & $\checkmark$ &&& $\checkmark$& & $\checkmark$ \cr
& $7/6$ & $2$ & $3$ &&&& $\checkmark$ &$\checkmark$ &$\checkmark$ &&$\checkmark$ \cr
& $5/6$ & $2$ & $\bar3$ &&&&&&& $\checkmark$ & \cr
& $1/2$ & $2$ & $1$ &&&&&& $\checkmark$ & $\checkmark$ & \cr
& $1/3$ &  $3$ & $\bar3$ & $\checkmark$ & $\checkmark$ &&&&&& \cr \hline
\multirow{8}{1em}{$1$} & $0$ & $1$ & $1$ & $\checkmark$ && $\checkmark$ & $\checkmark$ & $\checkmark$ && & \cr
& $2/3$ & $1$ & $3$ & $\checkmark$ &&&&&& $\checkmark$  & \cr
& $5/3$ & $1$ & $3$ &&& $\checkmark$ &&&& & \cr
& $1/6$ & $2$ & $3$ &&&& $\checkmark$ &&& & \cr
& $5/6$ & $2$ & $\bar3$ &&&&& $\checkmark$ && & \cr
& $0$ & $3$ & $1$ && $\checkmark$ &&&&& & \cr
& $2/3$ & $3$ & $3$ & $\checkmark$ & $\checkmark$ &&&&& & \cr \hline
\multirow{4}{1em}{$1'$}& $1/3$ & $1$ & $\bar3$ &&&&&& $\checkmark$ && $\checkmark$  \cr
& $1/2$ & $2$ & $\bar3$ &&&&&&&&  $\checkmark$ \cr
& $5/6$ & $2$ & $\bar3$ &&&&&&& $\checkmark$ & \cr
& $7/6$ & $2$ & $3$ &&&&&& $\checkmark$ &&  $\checkmark$ \cr \hline
\end{tabular}
\end{table*}

\section{Underlying heavy new physics}

Having selected the effective operators relevant for the processes of interest, it is useful to derive the types of NP that can generate such interactions at tree level in order to understand what the limits (or, hopefully, the signals) from the FCC-ee represent physically. 

All operators are of the form $( \bar \psi_1 \Gamma_a \psi_2)(\bar\psi_3 \Gamma^a \psi_4)$, where the $ \Gamma_a \, (\Gamma^a) $ represent covariant (contravariant) Dirac matrices. This type of operator can be generated at tree level only through the exchange of a boson $X$. If the exchange is in the $1,2 \to 3,4 $ channel, $X$ will have the quantum numbers of the composite $( \bar \psi_1 \Gamma_a \psi_2)$. However, the exchange may also occur in the $1,3 \to 2,4 $ and $1,4 \to 2,3$ channels, for which the quantum numbers of $X$ can be  obtained using Fierz transformations. A limit on $ \Lambda_{\tt eff}$ then corresponds to a limit on $ M_X/\sqrt{g_X g_X'}$ where $M_X$ denotes the $X$ mass and $ g_X,\,g_X' $ its couplings to an appropriate SM fermion pair.

Using this, we find the list of possible heavy bosons $X$, which is given in Table \ref{tab:NP}.\footnote{We do not display the $O(1)$ numerical factors associated with the Fierz transformations.} We will not, however, attempt to construct realistic models that accommodate one or more of these bosons, as such a (lengthy) exercise lies outside the scope of this investigation. In particular, the operators generated by spin $1'$ boson exchanges are included for completeness (see table \ref{tab:NP} for notation), though 
interactions involving an antisymmetric tensor field are absent 
in underlying renromalizable 4-dimensional NP models. 

\section{Single top-quark production at the FCC-ee with $E_{CM}=240$ GeV \label{sec:FCC-ee240}}

As previously noted, we consider the future FCC-ee collider with 
$E_{CM}=240$ GeV and study the expected sensitivity of the $b_p$-odd signals arising from the single top-quark production processes $e^+ e^- \to t j$ and $e^+ e^- \to t j \gamma$ (+ c.c.), 
since the on-shell single top production modes $e^+e^- \to tjZ$ and $e^+e^- \to tjW$ are kinematically inaccessible 
at the FCC-ee with $E_{CM}=240$ GeV.  

The $tj$ and $tj \gamma$ single top channels lead   to the following $2j$ and $4j$ signals (after the top decays via $t \to bW(\to jj)$ or $t \to bW(\to \ell \nu)$, see Table \ref{tab:3}):
\begin{eqnarray}
    e^+ e^- \to 4j ~&,&~ e^+ e^- \to 4j + \gamma ~, \label{channels4j}\\
    e^+ e^- \to 2j + \ell + \missET ~&,& e^+ e^- \to ~2j + \ell + \gamma + \missET ~. \label{channels2j}
\end{eqnarray}

The total effective cross-section (i.e., $\bar\sigma_{1j_b}$ in  Eqs.~\ref{sigmabar}) for the $b_p$-odd single $b$-jet signals in the multi-jets channels of Eqs.~\ref{channels4j} and \ref{channels2j} can be written in general as:
\begin{eqnarray}
\bar\sigma_{1j_b}=\bar\sigma_{1j_b}^{SM} + \frac{1}{\Lambda_{\tt eff}^2} \cdot \bar\sigma_{1 j_b}^{INT}  + \frac{1}{\Lambda_{\tt eff}^4} \cdot \bar\sigma_{1j_b}^{NP} ~, \label{NP-CSX}
\end{eqnarray}
where $\bar\sigma_{1j_b}^{SM}$ is the SM contribution which is dominated by the $b_p$-even SM background to these multi-jet signals due to non-perfect $b$-tagging and purity. Furthermore, 
$\bar\sigma_{1j_b}^{NP}$ denotes the $b_P$-odd $\text{NP}^2$ term originating from the single top production processes in Eqs.~\ref{Eq:tj} and \ref{Eq:tjV}.
The corresponding interference term, $\bar\sigma_{1j_b}^{INT}/\Lambda_{\tt eff}^2$, is negligibly small since, beyond its $1/\Lambda_{\tt eff}^2$ suppression, this term is further reduced by the off-diagonal CKM elements $V_{cb}$ and/or $V_{ub}$ within the SM diagrams, as well as by factors of light-fermion masses. This last arises because the interference term is generated via SM interactions that explicitly break the global $U(3)^5$ symmetry of the SM when Yukawa and CKM couplings are ignored (see \cite{our_EIC_paper}). 

Note also that the irreducible $b_P$-odd SM contribution to $\bar\sigma_{1j_b}^{SM}$, is generated only by diagrams with an insertion 
of the 3rd generation off-diagonal CKM factors, so that this contribution is also 
highly suppressed by $|V_{cb}|^2$ and $|V_{ub}|^2$ and is therefore neglected; we find that this irreducible contribution to $\bar\sigma_{1j_b}^{SM}$ is at best two orders of magnitude smaller than the $b_P$-even contributions to $\bar\sigma_{1j_b}^{SM}$ (i.e., from non-perfect $b$-tagging and purity).

Following Eqs.~\ref{sigmabar} and \ref{nml}, 
we can calculate the effective  single $b$-jet SM and NP cross-sections in the four channels listed in Eqs.~\ref{channels4j} and \ref{channels2j}, 
for the SM and for the three classes of $eetu$ and $eetc$ 4-Fermi operators in 
Table \ref{tab:3}, scalar ($S$), vector ($V$) and tensor ($T$): \\

\uuline{$e^+ e^- \to 4j,~ 4j + \gamma$ }
\begin{eqnarray}
   && \underline{NP(S,V,T) - eetu:} \cr 
    && \bar\sigma_{1j_b}^{NP} = P_0^1 \left[ P_1^1  P_0^2 \epsilon_b (1-t_c) (1-t_j)^2 + P_1^1  P_0^2 (1- \epsilon_b) t_c (1-t_j)^2  + P_0^1 P_1^2 (1- \epsilon_b) (1-t_c) t_j (1-t_j) \right] \sigma_{112}^{S,V,T} \cr
    && \qquad \qquad  + \left[ P_1^1 P_0^3  \epsilon_b (1-t_j)^3  + P_0^1 P_1^3 (1- \epsilon_b) t_j (1-t_j)^2 \right] \sigma_{103}^{S,V,T}   ~, \label{sigNP4jtu}
    \end{eqnarray}
    \begin{eqnarray}
    && \underline{NP(S,V,T) - eetc:} \cr
    && \bar\sigma_{1j_b}^{NP}= P_0^1 \left[ P_1^1  P_0^2 \epsilon_b (1-t_c) (1-t_j)^2 + P_1^1  P_0^2 (1- \epsilon_b) t_c (1-t_j)^2  + P_0^1 P_1^2 (1- \epsilon_b) (1-t_c) t_j (1-t_j) \right] \sigma_{112}^{S,V,T}  \cr 
    &&  \qquad \qquad + P_0^1 \left[ P_1^1  P_0^2 \epsilon_b (1- t_c)^2 (1-t_j)  + P_0^1 P_1^2 (1- \epsilon_b) t_c (1-t_c) (1-t_j) + P_1^1  P_0^2 (1- \epsilon_b) (1-t_c)^2 t_j \right] \sigma_{121}^{S,V,T}  ~, \label{sigNP4jtc}
\end{eqnarray}
\begin{eqnarray}
    && \underline{SM:} \nonumber \\
    && \bar\sigma_{1j_b}^{SM} = P_1^4 \left[ 
    \epsilon_b (1-\epsilon_b)^3 \sigma_{400}^{SM} +
    t_c (1-t_c)^3 \sigma_{040}^{SM} + t_j (1-t_j)^3 \sigma_{004}^{SM} \right]  \cr
    && \qquad \qquad +P_0^2 P_1^2 \left[ 
    \epsilon_b (1-\epsilon_b) (1-t_c)^2 + 
    t_c (1-t_c) (1-\epsilon_b)^2   
    \right] \sigma_{220}^{SM} \cr
    && \qquad \qquad +P_0^2 P_1^2 \left[ 
    \epsilon_b (1-\epsilon_b) (1-t_j)^2 + 
    t_j (1-t_j) (1-\epsilon_b)^2   
    \right] \sigma_{202}^{SM}  \cr
    && \qquad \qquad +P_0^2 P_1^2 \left[ 
    t_c (1-t_c) (1-t_j)^2 + 
    t_j (1-t_j) (1-t_c)^2   
    \right] \sigma_{022}^{SM} \cr
    && \qquad \qquad +\left[ P_1^1 P_0^3  
    t_c (1-t_j)^3 + P_0^1 P_1^3
     (1-t_c) t_j (1-t_j)^2 \right] \sigma_{013}^{SM} ~, \label{sigSM4j}
\end{eqnarray}
    
\uuline{$e^+ e^- \to 2j + \ell +\missET, ~ 2j + \ell + \gamma + \missET$}:

\begin{eqnarray}
    NP(S,V,T) - eetu: && \bar\sigma_{1j_b}^{NP} = P_0^1 P_1^1  \left[ \epsilon_b (1-t_j)  + (1- \epsilon_b) t_j \right] \sigma_{101}^{S,V,T} ~, \label{sigNP2jtu} \\
    NP(S,V,T) - eetc: && \bar\sigma_{1j_b}^{NP} = P_0^1 P_1^1  \left[ \epsilon_b (1-t_c)  + (1- \epsilon_b) t_c \right] \sigma_{110}^{S,V,T} ~. \label{sigNP2jtc}
\end{eqnarray}
\begin{eqnarray}
   SM: &&  \bar\sigma_{1j_b}^{SM} = P_1^2 t_j (1-t_j) \sigma_{002}^{SM}  + P_0^1 P_1^1 t_j (1-t_j) t_c (1-t_c) \sigma_{011}^{SM} ~, \label{sigSM2j}
\end{eqnarray}
where, for example, the $nml=202$ SM contribution to the $4j$ and $4j +\gamma$ final states is
\begin{eqnarray}
\sigma_{202}^{SM} &=& \sigma(e^+ e^- \to b \bar b gg) + \sigma(e^+ e^- \to b \bar b u \bar u) + \sigma(e^+ e^- \to b \bar b d \bar d) + 
\sigma(e^+ e^- \to b \bar b s \bar s) ~, 
\end{eqnarray}
while for any of the scalar, vector or tensor $eetu$ NP contributions to the $nml=112$ cross-section in the $4j$ and $4j +  \gamma$ channels we have
\begin{eqnarray}
\sigma_{112}^{S,V,T} &=& \sigma(e^+ e^- \to t \bar u \to 
b \bar u c \bar s) + \sigma(e^+ e^- \to \bar t u \to 
\bar b u \bar c s) + {\tt c.c.} ~, \label{CSX-example}
\end{eqnarray}
and similarly for all other SM and NP multi-jets cross-sections, $\sigma_{nml}^{SM}$ 
and $\sigma_{nml}^{S,V,T}$, that contribute to the overall single $b$-jet effective cross-sections $\bar\sigma_{1j_b}^{SM}$ and  $\bar\sigma_{1j_b}^{NP}$. 
\begin{figure}[htb]
  \centering
\includegraphics[width=0.45\textwidth]{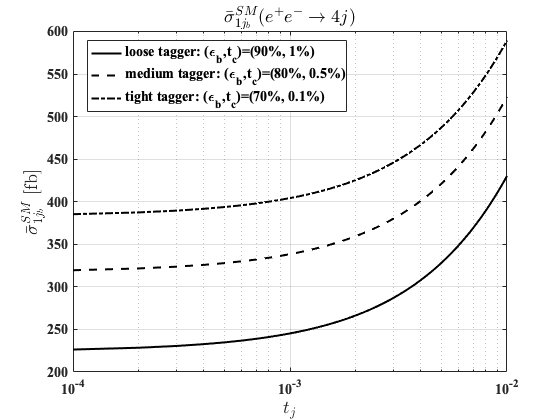}
\includegraphics[width=0.45\textwidth]{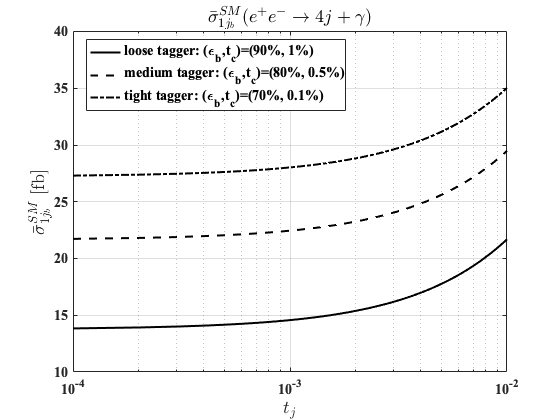}
\includegraphics[width=0.45\textwidth]{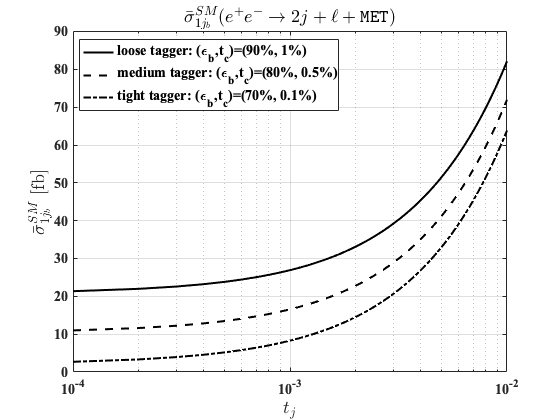}
\includegraphics[width=0.45\textwidth]{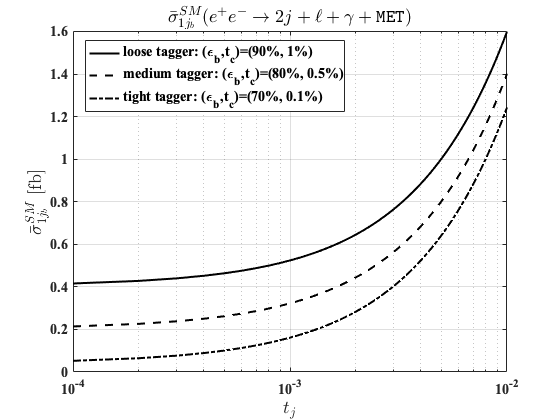}
\caption{The expected SM (background) effective single $b$-jet cross-sections 
 in the $4j$ (upper left), 
$4j + \gamma$ (upper right), $2j + \ell + \missET$ (lower left) and $2j + \ell + \gamma+ \missET$ (lower right) channels, as a function of the light-jet purity factor $t_j$, for 
the "loose", "medium" and "tight" tagger
scenarios, as indicated. See also text.}
  \label{fig:2}
\end{figure}

All cross-sections for the channels considered above, are calculated
using {\sc MadGraph5\_aMC@NLO}~\cite{madgraph5} at LO parton-level and with the SMEFTsim model of~\cite{SMEFTsim1,SMEFTsim2} for the EFT framework. For all the processes considered we apply acceptance cuts on the transverse momentum and pseudo-rapidity distributions on the jets and charged leptons (where relevant): $p_{T_j} > 10$ GeV, $|\eta_j| < 2$ and $p_{T_\ell} > 10$ GeV, $|\eta_\ell| < 2$, as well as a minimum angular separation criterion between any two jets of $\Delta R_{jj} > 0.4$.

%
%

In Fig.~\ref{fig:2} we plot the effective single $b$-jet SM cross-sections ($\bar\sigma_{1j_b}^{SM}$) in all four multi-jet channels of Eqs.~\ref{channels4j} and \ref{channels2j},
as a function of the light-quark/gluon jet purity parameter $t_j$ and for three sets of values for $(\epsilon_b,t_c)$: "loose": $(\epsilon_b,t_c)=(0.9,0.01)$,  "medium": $(\epsilon_b,t_c)=(0.8,0.005)$ and "tight": $(\epsilon_b,t_c)=(0.7,0.001)$. We recall that  $\bar\sigma_{1j_b}^{SM} \to 0$ in the limit $V_{ub},V_{cb} \to 0$ for the unrealistic case of a perfect $b$-tagging and purity of the $b$-jet samples, i.e., if $\epsilon_b =1$ and $t_j,t_c =0$. We see, however, that even with a relatively high $b$-tagging efficiency and purity performance, e.g., $\epsilon_b \geq 70\%$ and $t_j,t_c \sim 0.1\%$, one expects $10^4$ - $10^6$ single $b$-jet SM background events (depending on the channel) with an integrated luminosity of $10$ ab$^{-1}$. Notice also the different behavior of the $4j$ and $2j$ channels for the  tagger scenarios presented. In particular, while   $\bar\sigma_{1j_b}^{SM}(e^+ e^- \to 4j,4j+\gamma)$ is larger for the  tighter taggers, $\bar\sigma_{1j_b}^{SM}(e^+ e^- \to 2j+X,2j+\gamma+X)$  decreases as the tagger performance becomes tighter. As we will see below, this will determine the dependence of the sensitivity to the NP on the tagger scenarios in the different channels. 

Another important point is that the SM background/contribution to all the multi-jets processes above is dominated by the vector-boson pair production processes $e^+ e^- \to W^+ W^-,ZZ$ followed by their decays; as will be shown in the next section, this gives an extra handle for improving the sensitivity (signal-over-background) to the new flavor physics of our interest.



To assess the sensitivity of the $b_P$-odd 
NP signals of interest, we use (for each of the channels considered) the number of expected single $b$-jet events corresponding to the total effective cross-section $\bar\sigma_{1j_b}$ in Eq.\ref{NP-CSX}, which is a function of $\epsilon_b,t_c,t_j$ and the NP scale 
$\Lambda_{\tt eff}$:
\begin{eqnarray}
    N_{1j_b}(\epsilon_b,t_c,t_j,\Lambda_{\tt eff}) = {\cal L} \cdot {\cal A} \cdot \bar\sigma_{1j_b}(\epsilon_b,t_c,t_j,\Lambda_{\tt eff}) ~, \label{NCC} 
\end{eqnarray}
where we will henceforward take ${\cal L}=10$ ab$^{-1}$ as the projected total integrated luminosity for the FCC-ee in the $\sqrt{s}=240$ GeV running phase and ${\cal A}$ is the overall detector acceptance + efficiency factor which we will set below to ${\cal A}=0.8$.   

The sensitivity is then determined  
by comparing the theoretical shift due to the underlying 
$b_P$-odd interactions with the overall expected error ($\Delta$) 
in measuring the given quantity. 
Thus,
requiring a signal of at least $N_{SD}$ standard deviations, we have
\begin{eqnarray}
\left| N_{1j_b}(\epsilon_b,t_c,t_j,\Lambda_{\tt eff}) - N_{1j_b}(SM) \right| \geq N_{SD} \cdot \Delta \label{nsd}~,
\end{eqnarray}
where $N_{1j_b}(SM) = {\cal L} \cdot {\cal A} \cdot \bar\sigma_{1j_b}^{SM}$ is the expected number of single $b$-jet events in the SM (for any given channel), which also depends on the tagging and purity parameters $\epsilon_b,t_c,t_j$, see Eqs.\ref{sigSM4j} and \ref{sigSM2j}; recall that $\bar\sigma_{1j_b}^{SM}$ is dominated the $b_P$-even SM processes where light-jets are mis-identified as $b$-jets, i.e., by the non-perfect purity of the $b$-jet samples.
We include three contributions in the overall expected error $ \Delta $, 
\begin{center}
\begin{minipage}{0.5\textwidth}
\begin{description}
    \item[Statistical error] $\Delta_{\rm stat}=\sqrt{N_{1j_b}(\epsilon_b,t_c,t_j,\Lambda_{\tt eff})}$ 
\item[Systematic error] $\Delta_{\rm sys}=N_{1j_b}(\epsilon_b,t_c,t_j,\Lambda_{\tt eff}) \cdot \delta_s$ 
\item[Theory error] $\Delta_{\rm theor}= N_{1j_b}(\epsilon_b,t_c,t_j,\Lambda_{\tt eff})
\cdot \delta_t$
\end{description} 
\end{minipage}
\end{center}

\medskip

\noindent which we combine in
quadrature: $\Delta^2 = \Delta_{\rm stat}^2 + \Delta_{\rm sys}^2 + \Delta_{\rm theor}^2$, where $ \delta_{s}$ and $\delta_{t} $ denote the statistical and theoretical errors per event, respectively; $ \delta_s $ is usually estimated using experimental data from related processes and $ \delta_t $ is derived from the errors in the Monte Carlo integration used for calculating the various cross sections. We will always use below 
$ \delta_{s} = \delta_{t} = 0.01 $.  
\begin{figure}[htb]
  \centering
\includegraphics[width=0.45\textwidth]{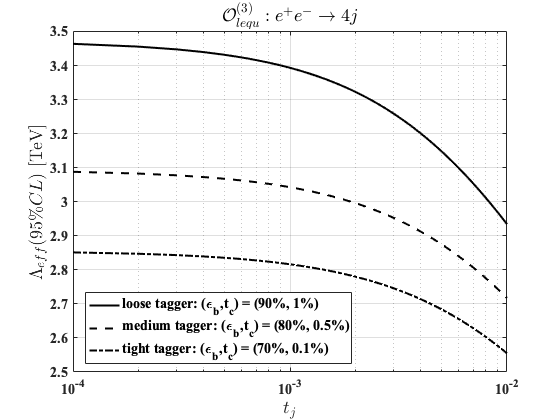}
\includegraphics[width=0.45\textwidth]{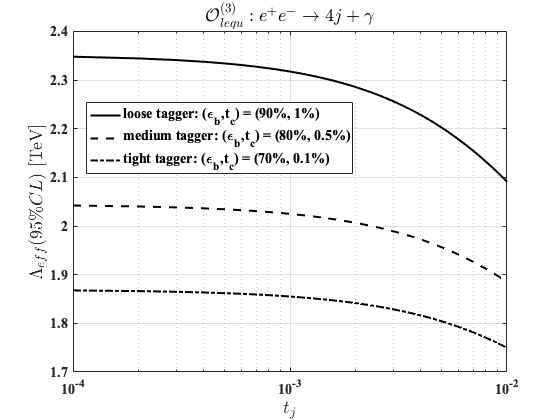}
\includegraphics[width=0.45\textwidth]{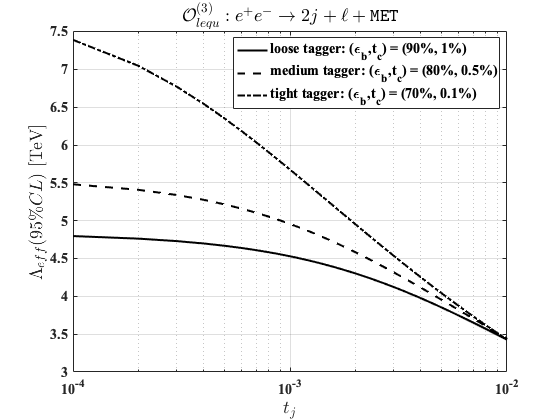}
\includegraphics[width=0.45\textwidth]{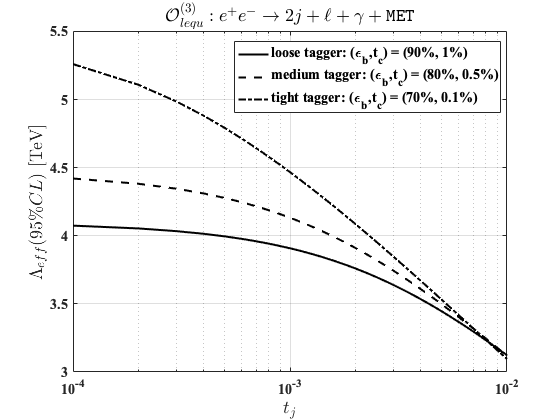}
\caption{Expected 95\% CL sensitivity to scale $\Lambda_{\tt eff}$ of the tensor operators $\op_{\ell e q u}^{(3)}(11k3)$ or $\op_{\ell e q u}^{(3)}(113k)$ ($k=1,2$ for the $eetu,eetc$ cases, respectively), as a function of the light-jet purity factor $t_j$, for 
the "loose", "medium" and "tight" $b$-tagging scenarios described in the text.
Results are shown in the $4j$ (upper left), $4j + \gamma$ (upper right), $2j + \ell + \missET$ (lower left), and $2j + \ell + \gamma+ \missET$ (lower right) channels. See also text.}
  \label{fig:3}
\end{figure}

In Fig.~\ref{fig:3} we plot the expected 95\% CL sensitivity to the scale $\Lambda_{\tt eff}$ of the tensor operators, $\op_{\ell e q u}^{(3)}(11k3)$ or $\op_{\ell e q u}^{(3)}(113k)$,   
that can be obtained by the simple $b$-jet counting method in all four channels: $e^+ e^- \to 4j$, $e^+ e^- \to 2j + \ell + \missET$, $e^+ e^- \to 4j + \gamma$ and $e^+ e^- \to 2j + \ell + \gamma + \missET$. 
Results are shown for the FCC-ee with $\sqrt{s}=240$ GeV and an integrated luminosity of ${\cal L}=10$ ab$^{-1}$. We see that an appreciably better sensitivity to the flavor-changing NP is expected in the two jets channels $e^+ e^- \to 2j + \ell + \missET$ and $e^+ e^- \to 2j + \ell + \gamma + \missET$, where the tight tagger performance is preferable in these channels, due to the behavior of the SM background under the purity factors, as explained above; in the $4j$ channels the signal-to-background ratio, $N_{1j_b}(\epsilon_b,t_c,t_j,\Lambda_{\tt eff})/N_{1j_b}(SM)$,  
decreases as the tagger performance becomes tighter, 
whereas in the $2j$ channels it increases with tighter tagger requirements.
\begin{table*}[htb]
\caption{Expected 95\% CL sensitivity to the NP scale $\Lambda_{\tt eff}$ of any one of the vector operators [$\op_{\ell q}^{(1)}(113k)$,$\op_{\ell q}^{(3)}(113k)$,$\op_{eu}(113k)$,$\op_{\ell u}(113k)$,$\op_{qe}(113k)$], 
the scalar operators [$\op_{\ell e q u}^{(1)}(113k)$,$\op_{\ell e q u}^{(1)}(11k3)$] and the tensor operators [$\op_{\ell e q u}^{(3)}(113k)$,$\op_{\ell e q u}^{(3)}(11k3)$], where $k=1$ and $k=2$ for the $eetu$ and $eetc$ contact terms (see also Tables \ref{tab:1} and \ref{tab:2}).
The reported  sensitivities  
are obtained by the $b$-jet counting method in the $2j$ channel 
$e^+ e^- \to tj \to 2j + \ell + \missET$ and are shown for the loose: $(\epsilon_b,t_c,t_j)=(90\%,1\%,0.1\%)$, medium: $(\epsilon_b,t_c,t_j)=(80\%,0.5\%,0.1\%)$ and tight: $(\epsilon_b,t_c,t_j)=(70\%,0.1\%,0.01\%)$ $b$-tagging scenarios. See also text. \label{tab:4}}
\renewcommand{\arraystretch}{1.5}
\begin{tabular}[t]{c|c|c|c}
 \multicolumn{4}{c}{95\% CL bound on $\Lambda_{\tt eff}$ [TeV] in the $2j+\ell+\missET$ channel} \\
 \hline 
 \hline
   & loose tagger &  medium tagger  & tight tagger\\
\hline
vector operators & 2.9   & 3.1  & 4.7 \\
scalar operators & 2.5   & 2.7  & 4.1 \\
tensor operators &  4.6 & 5.0  & 7.5 
\end{tabular}
\end{table*}

We reiterate that the sensitivity to the $eetu$ and $eetc$ contact terms, whether scalar, vector, or tensor, is nearly identical in each of the four channels. This is because the single-b-jet NP signals induced by these operators arise from the single-top production processes $e^+e^- \to t \bar u + \bar t u$ and $e^+e^- \to t \bar c + \bar t c$, which have the same cross-sections when $\Lambda_{\tt eff}(eetu) = \Lambda_{\tt eff}(eetc) $, up to terms proportional to $m_u/m_c$. Moreover, the 
effective single $b$-jet NP cross-sections in these $tu$ and $tc$ production channels, $\bar\sigma_{1 j_b}^{NP}(e^+e^- \to t \bar u + \bar t u)$ and $\bar\sigma_{1 j_b}^{NP}(e^+e^- \to t \bar c + \bar t c)$, are dominated by the terms that are $\propto \epsilon_b$, see 
Eqs.~\ref{sigNP4jtu}-\ref{sigNP4jtc} and Eqs.~\ref{sigNP2jtu}-\ref{sigNP2jtc}; the small difference between them is due to $t_j \neq t_c$. 

In Table \ref{tab:4} we list the expected 95\% CL sensitivity to $\Lambda_{\tt eff}$ in the $2j$ channel 
$e^+ e^- \to tj \to 2j + \ell + \missET$, at the FCC-ee with $\sqrt{s}=240$ GeV and an integrated luminosity of ${\cal L}=10$ ab$^{-1}$, for any one of the vector, scalar and tensor operators in Table \ref{tab:3}. 
In particular, the operators [$\op_{\ell q}^{(1)}(113k)$,$\op_{\ell q}^{(3)}(113k)$,$\op_{eu}(113k)$,$\op_{\ell u}(113k)$,$\op_{qe}(113k)$], [$\op_{\ell e q u}^{(1)}(113k)$,$\op_{\ell e q u}^{(1)}(11k3)$]  and [$\op_{\ell e q u}^{(3)}(113k)$,$\op_{\ell e q u}^{(3)}(11k3)$] are grouped in Table \ref{tab:4} as 
vector, scalar and tensor operators, respectively, since the same sensitivity, up to $\Delta \Lambda_{\tt eff} \sim \pm 0.1$ TeV, is found for any of the operators in each class; the small difference is a result of additional sub-leading contributions to the single $b$-jet signals from $eebu/eebc$ and $e \nu bu/e \nu bc$ interactions that are generated by some of these operators due to SU(2) invariance, e.g., 
$\op_{\ell e q u}^{(3)}(11k3)$ generates only the $eetu/eetc$ contact terms, while $\op_{\ell e q u}^{(3)}(113k)$ generates both the $eetu/eetc$ and the $e \nu bu/e \nu bc$ ones (see also discussion in section \ref{sec:3rdflavor}). 
We see that the expected 
sensitivity to these 4-fermion FC single top interactions at the FCC-ee, ranges from $\sim 2.5$ TeV to $\sim 7.5$ TeV, depending on 
the type of underlying physics and the tagger performance.

\begin{figure}[htb]
\centering
\includegraphics[width=0.6\textwidth]{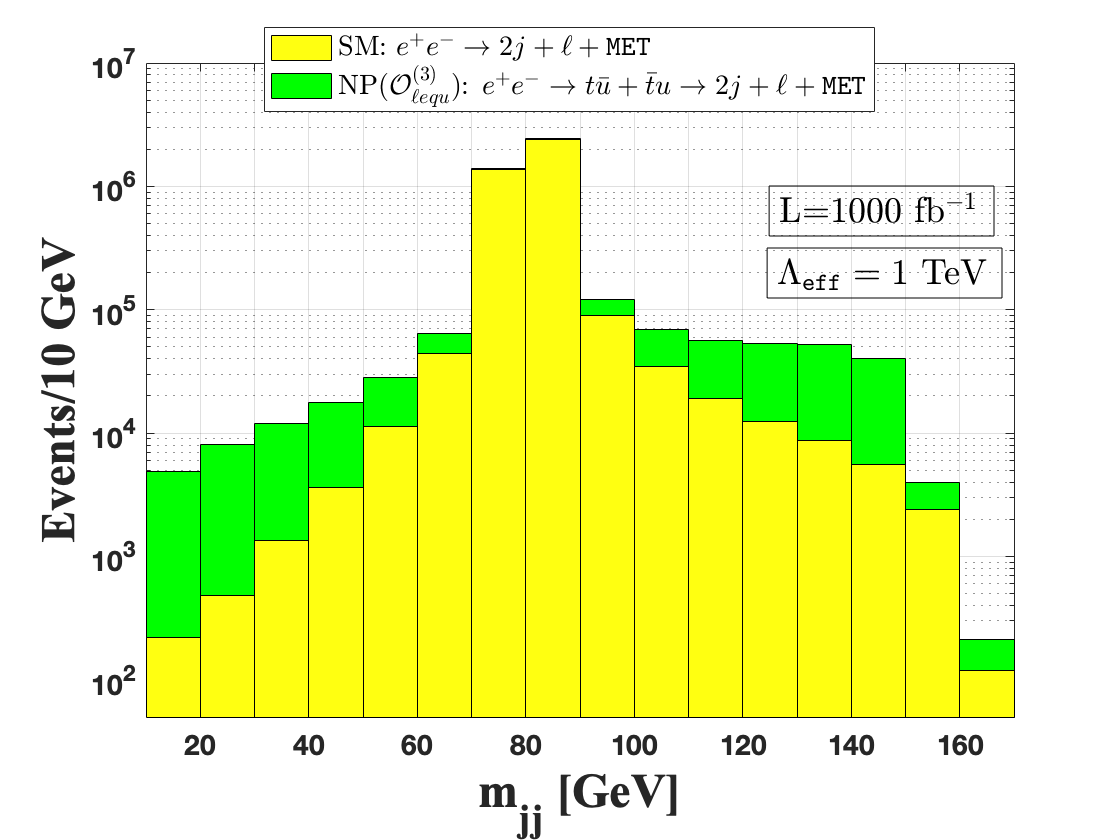}
\caption{Di-jet invariant mass distribution (stacked) for the SM and NP (for the tensor operator ${\cal O}_{lequ}^{(3)}$ case) contributions to the $2j$ signal 
$e^+ e^- \to 2j + \ell + \missET$. The distributions are shown per integrated luminosity of ${\cal L}=1000$~fb$^{-1}$ and for $\Lambda_{\tt eff}=1$ TeV.}
\centering
\label{fig:mjjdist}
\end{figure}

As noted above, the multi-jets SM  background processes to the $b_P$-odd (single $b$-jet) signals in the 
$e^+ e^- \to 2j +X $ and $e^+ e^- \to 4j +X $ channels that we consider, 
are dominated by the $WW$ and $ZZ$ production processes 
$e^+ e^- \to W^+ W^-,ZZ$, followed by the $W$ and $Z$ decays. For example, the SM contributions to the $b_P$-odd signals in the $e^+ e^- \to 2j + \ell + \missET$ channel (i.e., due to non-perfect purity) are $\sigma_{002}^{SM}$ and $\sigma_{011}^{SM}$ in Eq.~\ref{sigSM2j}, which are dominated by $e^+ e^- \to W^+(\to jj) W^-(\to \ell \nu_\ell)$ + c.c.. Therefore, the di-jet invariant mass distribution, $m_{jj}$, of the SM background to $e^+ e^- \to 2j + \ell + \missET$ picks at the $W$-boson threshold in this channel, while the genuine $b_P$-odd NP single top signal grows with $m_{jj}$.
This can be seen in Fig.~\ref{fig:mjjdist}, where we plot 
the $m_{jj}$ distribution in the $e^+ e^- \to 2j + \ell + \missET$ channel for the SM and the NP (from the tensor 4-Fermi operator) contributions. 

The distinct $m_{jj}$ behavior of the SM background and the single-top production processes induced by the $eetu/eetc$ contact interactions appears in all four channels considered here, namely those in Eqs.~\ref{channels4j} and \ref{channels2j}. Therefore, this difference in the $m_{jj}$ dependence of the NP and SM cross sections can be exploited as an additional discriminating variable to improve the signal-to-background ratio, and hence the sensitivity to the $b_P$-odd single-top NP signals. This point will be discussed in the next section.

\section{Enhanced sensitivity for single top kinematics  \label{sec:FCC-ee240_Mjj}}

\begin{figure}[htb]
\centering
\includegraphics[width=0.6\textwidth]{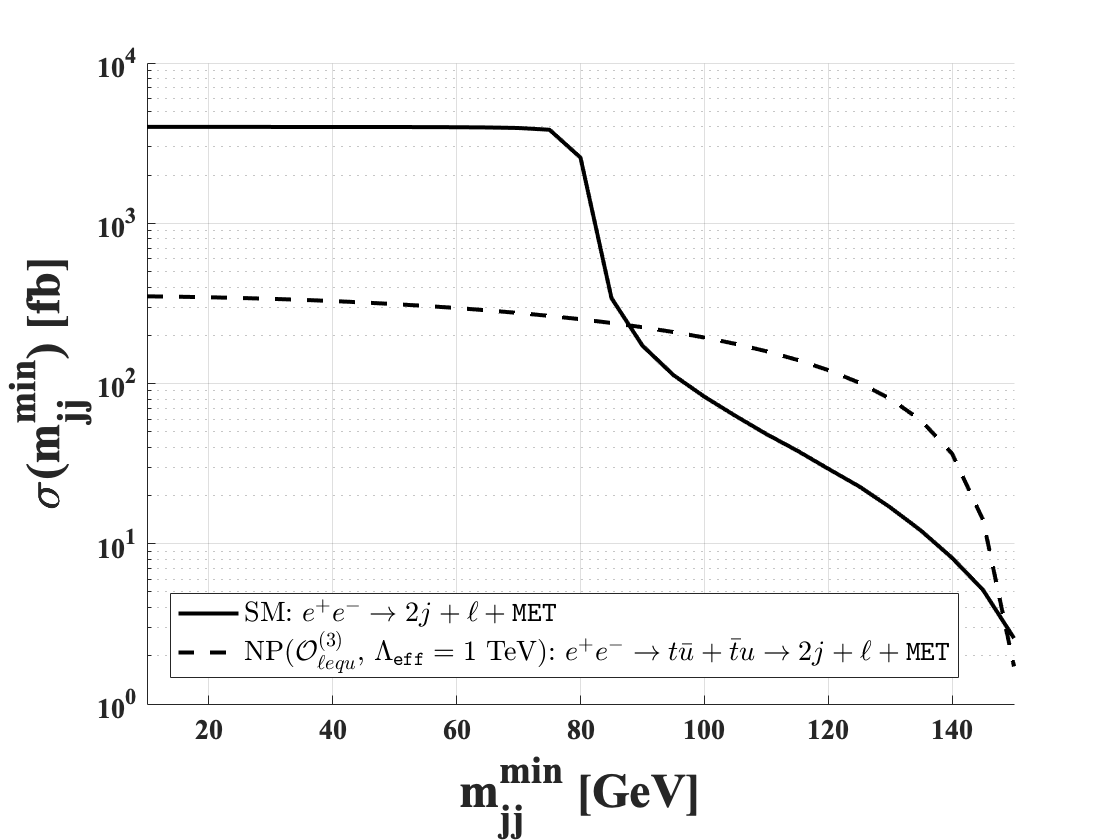}
\caption{The $m_{jj}^{\tt min}$ dependence of the SM and NP  cross-sections, $\sigma(m_{jj}^{\tt min})$, in the 
$e^+ e^- \to 2j + \ell + \missET$ channel. The NP cross-section corresponds to the tensor operator ${\cal O}_{lequ}^{(3)}$ with $\Lambda_{\tt eff}=1$ TeV. See also text.}
\centering
\label{fig:CSXmjj}
\end{figure}

Let us define the $m_{jj}$-dependent integrated cross-section, selecting events above a minimum value of $m_{jj}$:
\begin{eqnarray}
\sigma(m_{jj}^{\tt min}) \equiv 
\sigma( m_{jj} \geq m_{jj}^{\tt min}) = 
\int_{m_{jj} \geq m_{jj}^{\tt min}} d m_{jj} \frac{d\sigma}{dm_{jj}} ~, \label{CCSX}
\end{eqnarray}
where  $m_{jj}^{\tt min}$ can be chosen to optimize the sensitivity to the single top processes of interest. 
The dependence of the SM and NP cross sections on this cut is shown in Fig.~\ref{fig:CSXmjj}, where we plot $\sigma(m_{jj}^{\tt min})$ for  the $e^+ e^- \to 2j + \ell + \missET$ channel for both the SM and NP parts and 
the NP contribution is again assumed to originate from the $eetu$ tensor operator. Indeed, as as explained in the previous section, 
the SM background, $\sigma^{\rm SM}(m_{jj}^{\tt min})$, exhibits a sharp drop around the $W$-boson mass at $m_{jj} \sim 80$ GeV.

\begin{figure}[htb]
  \centering
\includegraphics[width=0.32\textwidth]{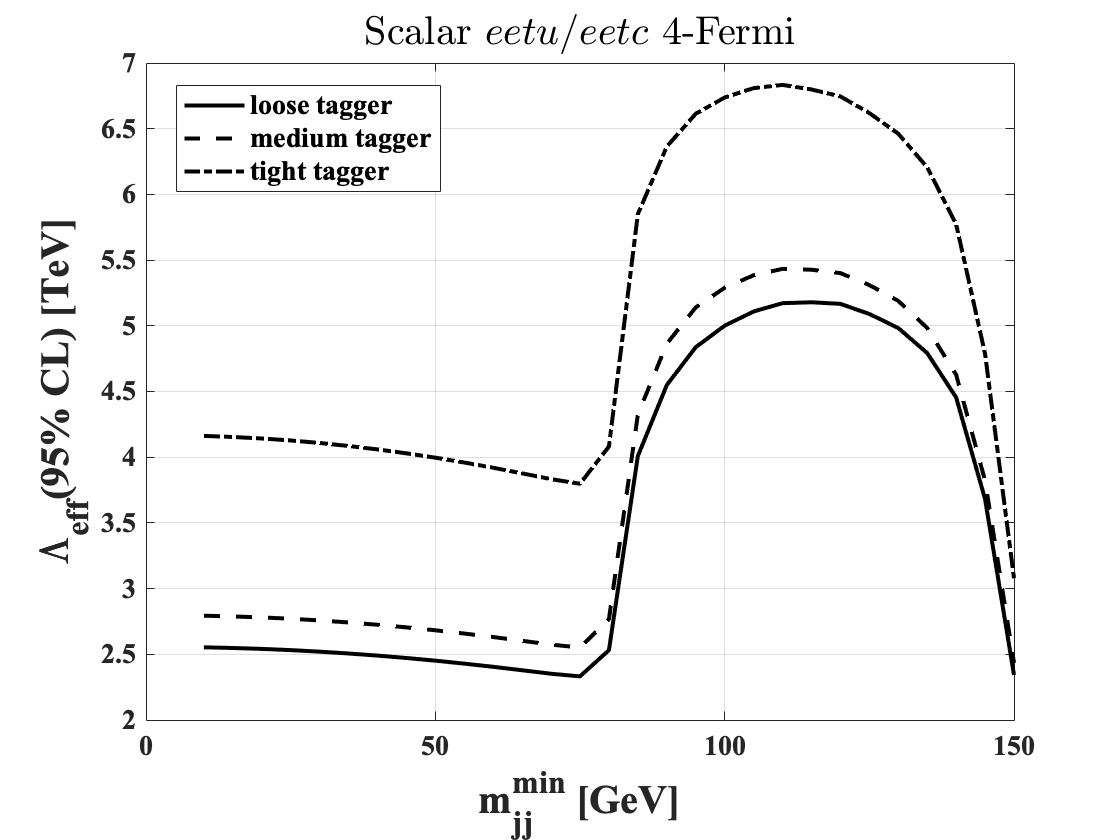}
\includegraphics[width=0.32\textwidth]{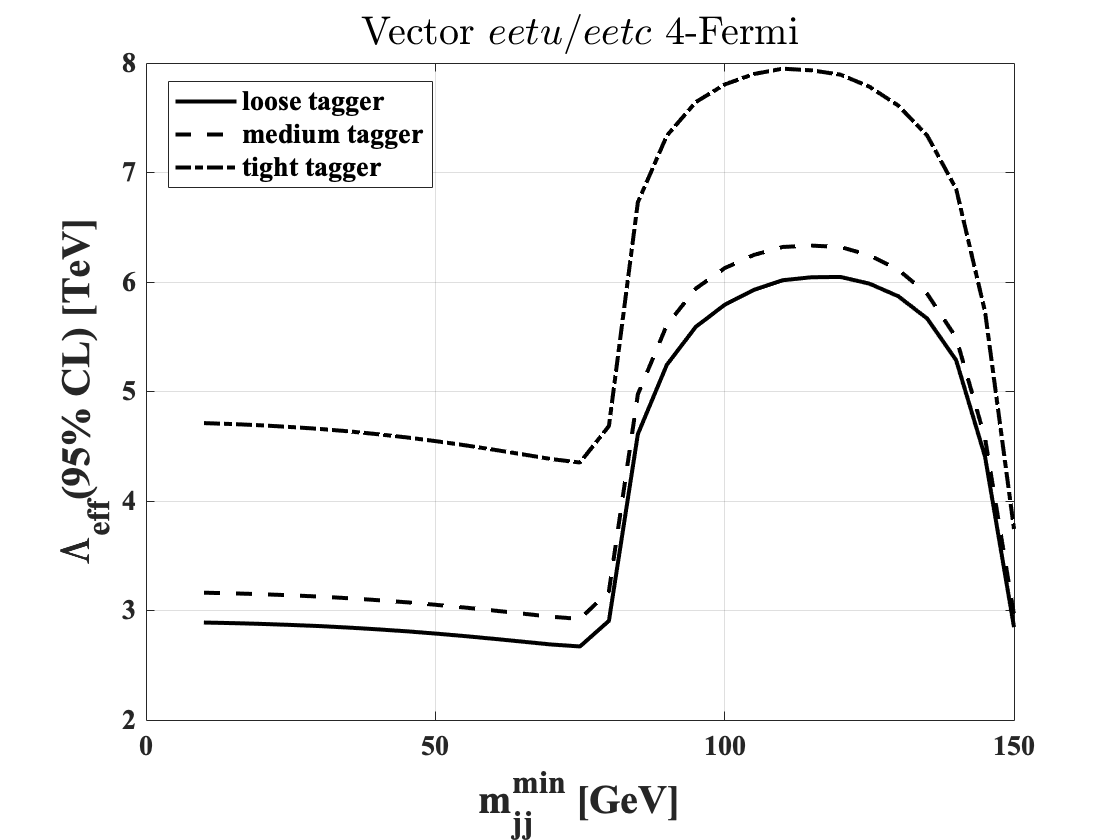}
\includegraphics[width=0.32\textwidth]{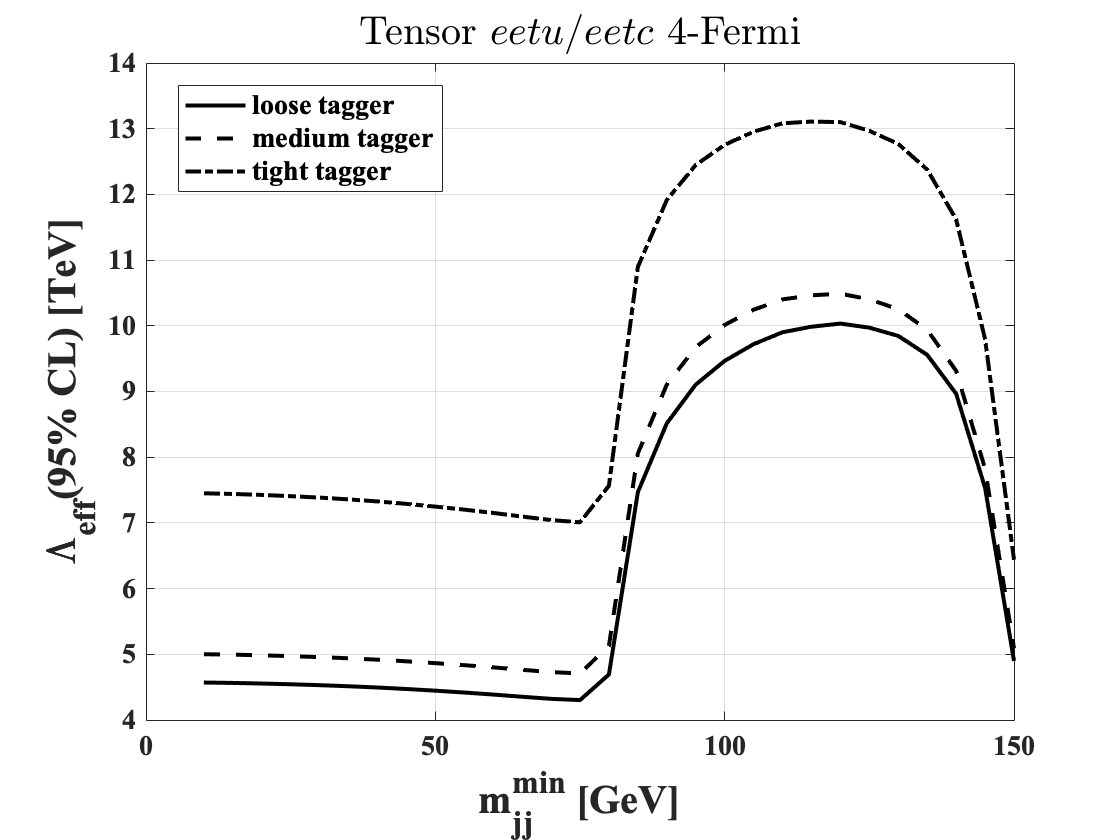}
\caption{Expected 95\% CL sensitivity to $\Lambda_{\tt eff}$ [TeV] from 
single top-quark production in the 
$2j + \ell + {\rm MET}$ channel, as a function of the 
lower cut on the di-jet invariant mass, $m_{jj}^{\tt min}$, for 
the "loose", "medium" and "tight" $b$-tagging scenarios described in the text. Results are shown for tensor (right), vector (middle) and scalar (left) $eetu/eetc$ operators, See also text.}
  \label{fig:mjj_sen}
\end{figure}

In Fig.~\ref{fig:mjj_sen} we show the 95\% CL sensitivity to the scale of the new flavor physics 
$\Lambda_{\tt eff}$ in the cases of scalar, vector and tensor $eetu$ and/or $eetc$ four-Fermi contact interactions.
As discussed in the previous section, approximately the same sensitivity is expected for the $eetu$ and $eetc$ operators. The sensitivity is shown as a function of $m_{jj}^{\tt min}$ for the $2j$ signal channel $e^+ e^- \to 2j + \ell + \missET$ for the loose, medium and tight tagger scenarios 
$(\epsilon_b,t_cj) = (90\%, 1\%, 0.1\%)$, 
$(\epsilon_b,t_cj) = (80\%, 0.5\%, 0.1\%)$ 
and 
$(\epsilon_b,t_cj) = (70\%, 0.1\%, 0.01\%)$, respectively. 
Evidently, a search for a $b_P$-odd signal from single top-quark production in the di-jet channel $e^+ e^- \to 2j + \ell + \missET$ would be sensitive to new FC scalar, vector and tensor $eetu$ and $eetc$ contact terms at the level of $\Lambda_{\tt eff} \sim 7,8$ and $13$ TeV, respectively, when 
a selection cut  of $m_{jj}^{\tt min} \sim 100$ GeV is applied on the di-jet events and with the tight tagger performance for the $b$-jet sample.
Notably, this corresponds to a sensitivity of $\Lambda_{\tt eff} \sim (30-50) \times E_{CM}$, namely 30 - 50 times the assumed CM energy of the FCC-ee, depending on the type of NP. These results illustrate the enormous usefulness the large luminosity and flavor identification efficiency expected in this collider will have as a handle in probing NP effects.

Finally, we note that if sizable $eetu$ and/or $eetc$ contact interactions of any type are present at multi-TeV energies, they would affect all four multi-jet channels considered here, namely those in Eqs.~\ref{channels4j} and \ref{channels2j}. Thus, the correlations among the excess events in these four channels could be further exploited to improve the sensitivity to this class of FC NP in top-quark interactions, as demonstrated in \cite{our_LFU_paper}.

\section{Summary}

We have explored the potential of a future FCC-ee, running at a center-of-mass energy of $E_{CM}=240$ GeV, to probe new flavor-changing interactions in the top-quark sector in the form of scalar, vector and tensor $eetu$ and $eetc$ 4-Fermi contact terms, which may be generated in the underlying theory by tree-level exchanges of multi-TeV new heavy states. 

In particular, we focused on the single top-quark signals: $e^+ e^- \to tj,~tj\gamma$ (+ c.c.), where $j$ is a light-quark or gluon jet (i.e., not a $b$-jet), which can be mediated by flavor-changing $eetu$ and $eetc$ 4-Fermi interactions.
These reactions lead (after the top-quark decays) to the multi-jets events $e^+ e^- \to 2j +X , ~ 4j+X$ ($X \in \ell^\pm, \missET, \gamma$), which contain a single $b$-jet from the top decay $t \to bW$. They are, therefore, odd under $b$-Parity, defined by $b_P = (-1)^{n}$ \cite{our_bP_paper}, where $n$ is the number of b-quarks jets in the final state in processes of the type $e^+ + e^- \to n \cdot j_b + m \cdot j_\ell + X$, where $n$($m$) is the number of produced $b$-quark(light-quark: $u,d,c,s$ and/or gluon) jets.

Thus, following the approach proposed in \cite{our_bP_paper}, we have shown that a search for these $b_P$-odd signals at the FCC-ee — based simply on counting the number of $b$-jets in the final state — can yield a sensitivity to the scalar, vector and tensor $eetu$ and $eetc$ contact interactions of $\Lambda \sim 10 \pm 3 $~TeV, depending on the type of operator and on the achievable $b$-tagging and purity performance. Remarkably, this reach corresponds to $\sim 30-50 \times E_{\text{CM}}$, improving upon current bounds on these flavor-changing interactions by approximately an order of magnitude.

This approach is viable because SM multi-jet production at the FCC-ee is dominated by final states with an even number of $b$-jets. In particular, as established in \cite{our_bP_paper}, the SM is $b_P$-even to high precision, as $b_P$-odd SM signals are suppressed by the small off-diagonal CKM matrix elements $|V_{ub}|^2$ and $|V_{cb}|^2$, rendering them unobservably small. More formally, in the limit $V_{j3}, V_{3j} \to 0$ ($j \neq 3$), the SM exhibits a global $U(1)_b$ "bottomness" symmetry that is exact to all orders of perturbation theory. Consequently, the only significant SM background to $b_P$-violating signals, such as the $b_P$-odd multi-jet events from single top production, arises from the mis-identification of light-quark and/or gluon jets as $b$-jets, 
i.e., from non-ideal $b$-tagging and purity of the $b$-jet sample. We have demonstrated that with a high flavor tagger performance, this background can be suppressed sufficiently to maintain the FCC-ee’s high sensitivity to NP.

Finally, we have mapped additional potential $b_P$-odd signals with high jet multiplicities — specifically $2j$, $4j$, and $6j$ final states containing one or three $b$-jets. These signatures can also be generated by the scalar, vector, and tensor $eetu$ and $eetc$ contact interactions via single top + $W/Z$ associated production, $e^+ e^- \to tjW, tjZ$. Such processes become accessible at the FCC-ee when operating at higher center-of-mass energies, such as the $t\bar{t}$ threshold phase with $E_{\text{CM}} \gtrsim 2m_t$.



\bibliographystyle{hunsrt.bst}
\bibliography{mybib2}

\end{document}